\documentclass{fac}

\usepackage{graphicx}
\usepackage{amssymb}
\usepackage{amsfonts}
\usepackage{amsmath}
\usepackage{dsfont}
\usepackage{MnSymbol}
\usepackage{mathtools}
\usepackage{epsfig}
\usepackage{epstopdf}

\newtheorem{theorem}{Theorem}[section]
\newtheorem{definition}[theorem]{Definition}

\newtheorem{lemma}[theorem]{Lemma}

\title[Draft of An Algebra of Actors Based on True Concurrency]
      {An Algebra of Actors Based on True Concurrency}

\author[Yong Wang]
    {Yong Wang\\
     College of Computer Science and Technology,\\
     Faculty of Information Technology,\\
     Beijing University of Technology, Beijing, China\\
     }

\correspond{Yong Wang, Pingleyuan 100, Chaoyang District, Beijing, China.
            e-mail: wangy@bjut.edu.cn}

\pubyear{2017}
\pagerange{\pageref{firstpage}--\pageref{lastpage}}

\begin{document}
\label{firstpage}

\makecorrespond

\maketitle

\begin{abstract}
An algebra of actors $\textrm{A}\pi$ fully captures the properties of actors based on asynchronous $\pi$-calculus, but, it is based on the interleaving bisimulation semantics. We adjust $\textrm{A}\pi$ to $\textrm{A}\pi_{tc}$ to make $\textrm{A}\pi$ having a truly concurrent semantics. We give the syntax and operational semantics of $\textrm{A}\pi_{tc}$, and also the truly concurrent semantics model and algebraic laws of $\textrm{A}\pi_{tc}$.
\end{abstract}


\section{Introduction}\label{int}

There are lots of work on true concurrency, including structures for true concurrency \cite{ES1} \cite{ES2} \cite{CM}, truly concurrent bisimilarities such as pomset bisimilarity, step bisimilarity, history-preserving (hp-)bisimilarity and the finest hereditary history-preserving (hhp-)bisimilarity \cite{HHP1} \cite{HHP2}. And also several kinds of logics for true concurrency were presented, such as a logic with reverse modalities \cite{RL1} \cite{RL2}, SFL logic \cite{SFL}, a uniform logic for true concurrency \cite{LTC1} \cite{LTC2} and a logic for weakly true concurrency \cite{WTC}. We also done several work on process algebra for true concurrency, including a calculus for true concurrency CTC \cite{CTC}, algebraic laws for true concurrency APTC \cite{ATC} and a calculus of truly concurrent mobile processes $\pi_{tc}$.

On the other hand, the actor computational model is a well-known truly concurrent computational model \cite{Actor1} \cite{Actor2} \cite{Actor3} \cite{Actor4}. An algebra of actors $\textrm{A}\pi$ \cite{Actor4} fully captures the properties of actors based on asynchronous $\pi$-calculus \cite{PI1} \cite{PI2}, but, it is based on the interleaving bisimulation semantics. In this paper, we adjust $\textrm{A}\pi$ to $\textrm{A}\pi_{tc}$ to make $\textrm{A}\pi_{tc}$ having a truly concurrent semantics.

This paper is organized as follows. In section \ref{am}, we briefly introduce the actor computational model. Then we introduce the syntax and operational semantics of $\textrm{A}\pi_{tc}$ in section \ref{sos}. In section \ref{tc}, we make $\textrm{A}\pi_{tc}$ to have a truly concurrent semantics. Finally, in section \ref{con}, we conclude this paper. 

\section{Actor Model}\label{am}

An actor is a concurrent object that encapsulates a set of states, a control thread and a set of local computations. It has a unique mail address and maintains a mail box to accept messages sent by other actors. Actors do local computations by means of processing the messages stored in the mail box sequentially and block when their mail boxes are empty. During processing a message in mail box, an actor may perform three candidate actions:

\begin{enumerate}
  \item \textbf{Send} action sends messages asynchronously to other actors by their mail box addresses;
  \item \textbf{Create} action creates new actors with new behaviors;
  \item \textbf{Ready} action makes the actor ready to process the next message from the mail box or block if the mail box is empty.
\end{enumerate}

\begin{figure}
  \centering
  %\vspace{5cm}
  \includegraphics{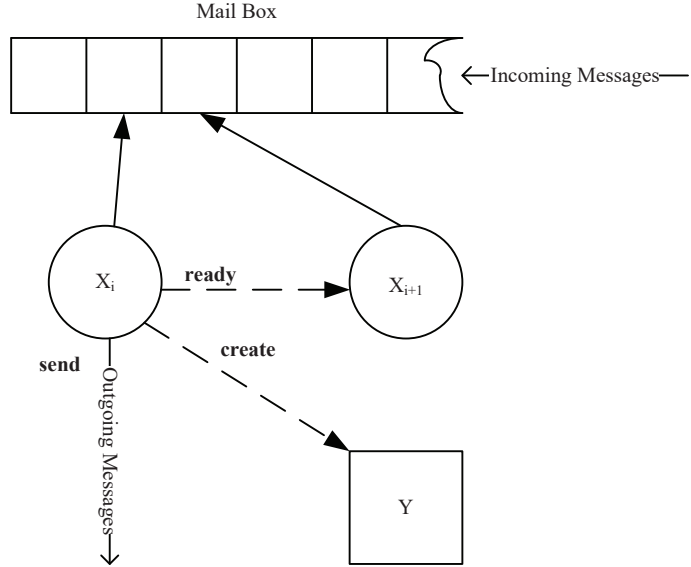}
  \caption{The actor computational model.}
  \label{Actor}
\end{figure}

The illustration of an actor model as shows in Fig.\ref{Actor} which is first shown in \cite{Actor1}.

Actors have the following three properties:

\begin{enumerate}
  \item \textbf{Uniqueness}: each actor has a unique name;
  \item \textbf{Freshness}: actors cannot be created with well-known names or names received in a message;
  \item \textbf{Persistence}: actors are persistent, they do not disappear after processing a message.
\end{enumerate}

Professor Gul Agha have done significant work on actors \cite{Actor2} \cite{Actor3} \cite{Actor4}, in which, $\textrm{A}\pi$ \cite{Actor4} is an algebra of actors that implements the above full properties of actors based on $\pi$-calculus \cite{PI1} \cite{PI2}. 

\section{Syntax and Operational Semantics}\label{sos}

We assume an infinite set $\mathcal{N}$ of (action or event) names, and use $a,b,c,\cdots$ to range over $\mathcal{N}$, use $x,y,z,w,u,v$ as meta-variables over names. We denote by $\overline{\mathcal{N}}$ the set of co-names and let $\overline{a},\overline{b},\overline{c},\cdots$ range over $\overline{\mathcal{N}}$. Then we set $\mathcal{L}=\mathcal{N}\cup\overline{\mathcal{N}}$ as the set of labels, and use $l,\overline{l}$ to range over $\mathcal{L}$. We extend complementation to $\mathcal{L}$ such that $\overline{\overline{a}}=a$. Let $\tau$ denote the silent step (internal action or event) and define $Act=\mathcal{L}\cup\{\tau\}$ to be the set of actions, $\alpha,\beta$ range over $Act$. And $K,L$ are used to stand for subsets of $\mathcal{L}$ and $\overline{L}$ is used for the set of complements of labels in $L$.

We write $\mathcal{P}$ for the set of configurations. Let $\widetilde{x}=x_1,\cdots,x_{ar(A)}$ and $\widetilde{y}=y_1,\cdots,y_{ar(A)}$ be tuples of distinct name variables, then $B\langle\widetilde{x};\widetilde{y}\rangle$ is called a configuration constant. Let the variable $\hat{z}$ range over $\{\emptyset,\{z\}\}$. The symbol $\widetilde{x},\widetilde{y}$ denotes the result of appending $\widetilde{y}$ to $\widetilde{x}$. The symbol $\widetilde{x},\hat{z}$ means that $\widetilde{x},z$ if $\hat{z}=\{z\}$; $\widetilde{x}$ otherwise. While $(\nu\hat{z})P$ means that $(\nu z)P$ if $\hat{z}=\{z\}$; $P$ otherwise. The symbol $\equiv_{\alpha}$ denotes equality under standard alpha-convertibility, note that the subscript $\alpha$ has no relation to the action $\alpha$.

Following $\textrm{A}\pi$ \cite{Actor4}, we retype the syntax and semantics of $\textrm{A}\pi_{tc}$, and adjust them to be suitable for true concurrency as follows.

\subsection{Syntax}

We give the syntax of $\textrm{A}\pi_{tc}$ as follows.

\begin{definition}[Syntax]\label{syntax}
A truly concurrent configuration $P$ in $\textrm{A}\pi_{tc}$ is defined inductively by the following formation rules:

\begin{enumerate}
  \item $B\langle\widetilde{x};\widetilde{y}\rangle\in\mathcal{P}$;
  \item $\textbf{0}\in\mathcal{P}$;
%  \item if $P\in\mathcal{P}$, then the Prefix $\tau.P\in\mathcal{P}$, for $\tau\in Act$ is the silent action;
  \item the Output $\overline{x}y\in\mathcal{P}$, for $x,y\in Act$;
  \item if $P\in\mathcal{P}$, then the Input $x(y).P\in\mathcal{P}$, for $x,y\in Act$;
  \item if $P\in\mathcal{P}$, then the Restriction $(\nu x)P\in\mathcal{P}$, for $x\in Act$;
%  \item if $P\in\mathcal{P}$, then the Match $[x=y]P\in\mathcal{P}$, for $x,y\in Act$;
  \item if $P_1,\cdots,P_n\in\mathcal{P}$, then $\textrm{case } x \textrm{ of }(y_1:P_1,\cdots,y_n:P_n)\in\mathcal{P}$;
  \item if $P,Q\in\mathcal{P}$, then the Composition $P\mid Q\in\mathcal{P}$;
\end{enumerate}

The standard BNF grammar of syntax of $\textrm{A}\pi_{tc}$ can be summarized as follows:

$$P::=B\langle\widetilde{x};\widetilde{y}\rangle\quad|\quad\textbf{0}\quad|\quad \overline{x}y\quad |\quad x(y).P \quad|\quad (\nu x)P \quad | \quad \textrm{case } x \textrm{ of }(y_1:P_1,\cdots,y_n:P_n)\quad |\quad P\mid P.$$
\end{definition}

For each behavior instantiation $B\langle\widetilde{x};\widetilde{y}\rangle$, a defining equation of the form

$$B\langle\widetilde{x};\widetilde{y}\rangle\overset{\text{def}}{=}(\widetilde{x};\widetilde{y})x_1(z).P$$

is assumed, where $P$ is a configuration.

The intuitions of the above constructs for actors, please refer to $\textrm{A}\pi$ \cite{Actor4}, we do not explain any more.

\begin{definition}[Free variables]
The free names of a configuration $P$, $fn(P)$, are defined as follows.

\begin{enumerate}
  \item $fn(B\langle\widetilde{x};\widetilde{y}\rangle)\subseteq\{\widetilde{x}\}\cup\{\widetilde{y}\}$;
  \item $fn(\textbf{0})=\emptyset$;
%  \item $fn(\tau.P)=fn(P)$;
  \item $fn(\overline{x}y.P)=fn(P)\cup\{x\}\cup\{y\}$;
  \item $fn(x(y).P)=fn(P)\cup\{x\}-\{y\}$;
  \item $fn((\nu x)P)=fn(P)-\{x\}$;
%  \item $fn([x=y]P)=fn(P)$;
  \item $fn(\textrm{case } x \textrm{ of }(y_1:P_1,\cdots,y_n:P_n))=fn(P_1)\cup\cdots\cup fn(P_n)$;
  \item $fn(P_1\mid P_2)=fn(P_1)\cup fn(P_2)$.
\end{enumerate}
\end{definition}

\begin{definition}[Bound variables]
Let $n(P)$ be the names of a configuration $P$, then the bound names $bn(P)=n(P)-fn(P)$.
\end{definition}

In $\overline{x}y$, $x(y)$ and $\overline{x}(y)$, $x$ is called the subject, $y$ is called the object and it may be free or bound.

\begin{definition}[Substitutions]\label{subs}
A substitution is a function $\sigma:\mathcal{N}\rightarrow\mathcal{N}$. For $x_i\sigma=y_i$ with $1\leq i\leq n$, we write $\{y_1/x_1,\cdots,y_n/x_n\}$ or $\{\widetilde{y}/\widetilde{x}\}$ for $\sigma$. For a configuration $P\in\mathcal{P}$, $P\sigma$ is defined inductively as follows:
\begin{enumerate}
  \item if $P$ is a configuration constant $B\langle\widetilde{x};\widetilde{y}\rangle=B\langle x_1,\cdots,x_n;y_1,\cdots,y_m\rangle$, then $P\sigma=B\langle x_1\sigma,\cdots,x_n\sigma;y_1\sigma,\cdots,y_m\sigma\rangle$;
  \item if $P=\textbf{0}$, then $P\sigma=\textbf{0}$;
%  \item if $P=\tau.P'$, then $P\sigma=\tau.P'\sigma$;
  \item if $P=\overline{x}y.P'$, then $P\sigma=\overline{x\sigma}y\sigma.P'\sigma$;
  \item if $P=x(y).P'$, then $P\sigma=x\sigma(y).P'\sigma$;
  \item if $P=(\nu x)P'$, then $P\sigma=(\nu x\sigma)P'\sigma$;
%  \item if $P=[x=y]P'$, then $P\sigma=[x\sigma=y\sigma]P'\sigma$;
  \item if $P=\textrm{case } x \textrm{ of }(y_1:P_1,\cdots,y_n:P_n)$, then $P\sigma=\textrm{case } x\sigma \textrm{ of }(y_1\sigma:P_1\sigma,\cdots,y_n\sigma:P_n\sigma)$;
  \item if $P=P_1\mid P_2$, then $P\sigma=P_1\sigma \mid P_2\sigma$.
\end{enumerate}
\end{definition}

\subsection{Type System}

According to the actor model, as in $\textrm{A}\pi$, actor names have uniqueness and freshness properties, and the persistence property is relaxed by permitting a \emph{sink} behavior. To assure such properties, as in $\textrm{A}\pi$, a type system is presented as follows.

As in \cite{Actor4}, $\bot,*\notin \mathcal{N}$, for $X\subset\mathcal{N}$, $X^*=X\cup\{\bot,*\}$; for $f:X\rightarrow X^*$, $f^*:X^*\rightarrow X^*$ is defined as $f^*(x)=f(x)$ if $x\in X$, and $f^*(\bot)=f^*(*)=\bot$. $\rho;f\vdash P$ is a typing judgement, where $\rho$ is the receptionist set of $P$, and $f:\rho\rightarrow \rho^*$ is a temporary name mapping function that relates actors in $P$ to the temporary names they have currently assumed. $f$, $f^*$, $f_1\oplus f_2$, $f|\rho$, $ch(\widetilde{x})$ have the same definitions and properties as those in \cite{Actor4}.

\begin{definition}[Type system]
The type system is consist of type rules that is the same as $\textrm{A}\pi$, we retype them in Table \ref{TS}.

\begin{center}
    \begin{table}
        \[\textbf{NIL}\quad \emptyset;\{\}\vdash\textbf{0} \quad \textbf{MSG}\quad \emptyset;\{\}\vdash \overline{x}y\]

        \[\textbf{ACT}\quad \frac{\rho;f\vdash P}{\{x\}\cup\hat{z};ch(x,\hat{z})\vdash x(y).P}\textrm{ if }\rho-\{x\}=\hat{z}, y\notin\rho, f=ch(x,\hat{z})\textrm{ if }x\in\rho;f=ch(\epsilon,\hat{z})\textrm{ otherwise }\]

        \[\textbf{CASE}\quad \frac{\forall 1\leq i\leq n,\rho_i;f_i\vdash P_i}{(\cup_i\rho_i);(f_1\oplus\cdots\oplus f_n)\vdash \textrm{ case }x \textrm{ of }(y_1:P_1,\cdots,y_n:P_n)}\textrm{ if }f_i\textrm{ are mutually compatible}\]

        \[\textbf{COMP}\quad \frac{\rho_1;f_1\vdash P_1\quad \rho_2;f_2\vdash P_2}{\rho_1\cup\rho_2;f_1\oplus f_2\vdash P_1\mid P_2}\textrm{ if }\rho_1\cap\rho_2=\emptyset\]

        \[\textbf{RES}\quad \frac{\rho;f\vdash P}{\rho-\{x\};f|(\rho-\{x\})\vdash (\nu x)P}\]

        \[\textbf{INST}\quad \{\widetilde{x}\};ch(\widetilde{x})\vdash B\langle\widetilde{x};\widetilde{y}\rangle \textrm{ if }len(\widetilde{x})=2\textrm{ implies }x_1\neq x_2\]

        \caption{Type rules of $\textrm{A}\pi_{tc}$}
        \label{TS}
    \end{table}
\end{center}
\end{definition}

The following theorem still holds, we retype it from $\textrm{A}\pi$.

\begin{theorem}
If $\rho;f\vdash P$ then $\rho\subset fn(P)$, and for all $x,y\in\rho$, $f(x)\neq x$, $f^*(f(x))=\bot$, and $f(x)=f(y)\notin \{\bot,*\}$ implies $x=y$. Furthermore, if $\rho';f'\vdash P$ then $\rho=\rho'$ and $f=f'$.
\end{theorem}

Since arbitrary substitution $\sigma$ on a configuration $P$ may destroy the uniqueness, freshness or persistence properties and cause $P$ to be an invalid $\textrm{A}\pi_{tc}$ term, we often assume that $\sigma$ is an one-to-one mapping. The following lemma also holds as in $\textrm{A}\pi$.

The following lemma says that the type system respects $\equiv_{\alpha}$.

\begin{lemma}
If $\rho;f\vdash P$ and $\sigma$ is one-to-one on $\rho$, then $\sigma(\rho);f\sigma\vdash P\sigma$.
\end{lemma}

\subsection{Operational Semantics}

The operational semantics is defined by LTSs (labelled transition systems), and it is detailed by the following definition.

\begin{definition}[Semantics]\label{semantics}
The operational semantics of $\textrm{A}\pi_{tc}$ corresponding to the syntax in Definition \ref{syntax} is defined by a series of transition rules, they are shown in Table \ref{TR}. Note that, these rules are adjusted to a truly concurrent version.

\begin{center}
    \begin{table}
        \[\textbf{INP}\quad \frac{}{x(y).P\xrightarrow{xz}P\{z/y\}} \quad \textbf{OUT}\quad \frac{}{\overline{x}y.P\xrightarrow{\overline{x}y}P}\]

        \[\textbf{BINP}\quad \frac{P\xrightarrow{xy}P'}{P\xrightarrow{x(y)}P'}\quad (y\notin fn(P)) \quad \textbf{RES}\quad \frac{P\xrightarrow{\alpha}P'}{(\nu y)P\xrightarrow{\alpha}(\nu y)P'} \quad (y\notin n(\alpha))\]

        \[\textbf{OPEN}\quad \frac{P\xrightarrow{\overline{x}y}P'}{(\nu y)P\xrightarrow{\overline{x}(y)}P'}\quad (x\neq y)\]

        \[\textbf{PAR}_1\quad \frac{P\xrightarrow{\alpha}P'\quad Q\nrightarrow}{P\mid Q\xrightarrow{\alpha}P'\mid Q}\quad (bn(\alpha)\cap fn(Q)=\emptyset) \quad \textbf{PAR}_2\quad \frac{Q\xrightarrow{\alpha}Q'\quad P\nrightarrow}{P\mid Q\xrightarrow{\alpha}P\mid Q'}\quad (bn(\alpha)\cap fn(P)=\emptyset)\]

        \[\textbf{PAR}_3\quad \frac{P\xrightarrow{\alpha}P'\quad Q\xrightarrow{\beta}Q'}{P\mid Q\xrightarrow{\{\alpha,\beta\}}P'\mid Q'}\quad (\beta\neq\overline{\alpha}, bn(\alpha)\cap bn(\beta)=\emptyset, bn(\alpha)\cap fn(Q)=\emptyset,bn(\beta)\cap fn(P)=\emptyset)\]

        \[\textbf{PAR}_4\quad \frac{P\xrightarrow{x_1(z)}P'\quad Q\xrightarrow{x_2(z)}Q'}{P\mid Q\xrightarrow{\{x_1(w),x_2(w)\}}P'\{w/z\}\mid Q'\{w/z\}}\quad (w\notin fn((z)P)\cup fn((z)Q))\]

        \[\textbf{COM}\quad \frac{P\xrightarrow{\overline{x}y}P'\quad Q\xrightarrow{xy}Q'}{P\mid Q\xrightarrow{\tau}P'\mid Q'}\]

        \[\textbf{CLOSE}\quad \frac{P\xrightarrow{\overline{x}(y)}P'\quad Q\xrightarrow{xy}Q'}{P\mid Q\xrightarrow{\tau}(\nu y)(P'\mid Q')}\quad (y\notin fn(Q))\]

        \[\textbf{BEHV}\quad\frac{x_1(z).P\{(\widetilde{u},\widetilde{v})/(\widetilde{x},\widetilde{y})\}\xrightarrow{\alpha}P'}{ B\langle \widetilde{u};\widetilde{v}\rangle\xrightarrow{\alpha}P'}\quad (B\langle\widetilde{x};\widetilde{y}\rangle\overset{\text{def}}{=}(\widetilde{x};\widetilde{y})x_1(z).P)\]

        \[\textbf{BRNCH}\quad \frac{}{\textrm{case }x \textrm{ of }(y_1:P_1,\cdots,y_n:P_n)\xrightarrow{\tau}P_i}\quad (\textrm{if }x=y_i)\]

        \caption{Transition rules of $\textrm{A}\pi_{tc}$}
        \label{TR}
    \end{table}
\end{center}
\end{definition}

The intuitions of transition rules in Table \ref{TR} for the actor computational model are the same as those of $\textrm{A}\pi$, the differences are that the $\textbf{PAR}$ rule is replaced by four rules $\textbf{PAR}_1$--$\textbf{PAR}_4$. The rules $\textbf{PAR}_1$--$\textbf{PAR}_4$ capture the truly concurrent semantics.

The following theorem still hold for transition rules in Table \ref{TR}, which says the well-typed terms are closed under transitions.

\begin{theorem}
(1)If $P$ is well-typed and $P\xrightarrow{\alpha}P'$ then $P'$ is well-typed;

(2)If $P$ is well-typed and $P\xrightarrow{\{\alpha_1,\cdots,\alpha_n\}}P'$ then $P'$ is well-typed.
\end{theorem}

As in $\textrm{A}\pi$, not every trace produced by the transition system in Table \ref{TR} corresponds to an actor computation. We have the following instance,

$$(\nu x)(x(u).P\mid \overline{x}x\mid\overline{y}x)\xrightarrow{\{x(u),\overline{x}x,\overline{y}x\}}P$$

by the transition system in Table \ref{TR}. But the above transition does not correspond to an actor computation, since there cannot be an actor named $x$ in the environment. Similarly, we also need the notation of $\rho$-well-formed trace with $\rho$ as an initial receptionist set, we retype it and adjust it to truly concurrent semantics as follows.

\begin{definition}
For a set of names $\rho$ and trace $s$, $rcp(\rho,s)$ is inductively defined as follows:

\begin{enumerate}
  \item $rcp(\rho,\epsilon)=\rho$;
  \item $rcp(\rho,s.(\hat{y_1})\cdots(\hat{y_n})(x_1y_1\mid\cdots\mid x_ny_n))=rcp(\rho,s)$;
  \item $rcp(\rho,s.(\hat{y_1})\cdots(\hat{y_n})(\overline{x_1}y_1\mid\cdots\mid \overline{x_n}y_n))=rcp(\rho,s)\cup\hat{y_1}\cup\cdots\cup\hat{y_n}$.
\end{enumerate}

We say $s$ is $\rho$-well-formed if $s=s_1.(\hat{y_1})\cdots(\hat{y_n})(\overline{x_1}y_1)\mid\cdots\mid \overline{x_n}y_n.s_2$ implies $x_1\notin rcp(\rho,s_1),\cdots,x_n\notin rcp(\rho,s_1)$, and $s$ is well-formed if it is $\emptyset$-well-formed.
\end{definition}

Let $\Rightarrow$ denote the reflexive transitive closure of $\xrightarrow{\tau}$, $\xRightarrow{\alpha}$ denote $\Rightarrow\xrightarrow{\alpha}\Rightarrow$, and $P\xRightarrow{s}$ denote $P\xRightarrow{s}P'$ for some $P'$. Then, the following lemma still hold for truly concurrent semantics.

\begin{lemma}
Let $P\mid Q$ be a well-typed $\textrm{A}\pi_{tc}$ term with $rcp(P)=\rho_1$ and $rcp(Q)=\rho_2$. Then $P\mid Q\Rightarrow$ can be unzipped into $P\xRightarrow{s}$ and $Q\xRightarrow{\overline{s}}$ such that $s$ is $\rho_1$-well-formed and $\overline{s}$ is $\rho_2$-well-formed.
\end{lemma}

Similarly, in $\rho$-well-formed traces, we only consider the following traces $s$ such that if $s=s_1.(\alpha_1\mid\cdots\mid\alpha_n).s_2$, where $(\rho\cup n(s_1)\cup fn(\alpha_1)\cup\cdots\cup fn(\alpha_n)) \cap bn((\alpha_1\mid\cdots\mid\alpha_n).s_2)=\emptyset$.

The transition sequences are also further constrained by a fairness requirement. Different to $\textrm{A}\pi$, the following transition sequences are fair in $\textrm{A}\pi_{tc}$.

\begin{eqnarray}
Diverge\langle x\rangle\mid\overline{x}u\mid y(v).\overline{v}v\mid\overline{y}v&\Rightarrow & Diverge\langle x\rangle\mid\overline{x}u\mid\overline{v}v\nonumber\\
&\xRightarrow{\overline{v}v}& Diverge\langle x\rangle\mid \overline{x}u \nonumber\\
&\xrightarrow{\tau}&Diverge\langle x\rangle \mid \overline{x}u \nonumber\\
&\xrightarrow{\tau}&\cdots\nonumber
\end{eqnarray}

where $Diverge\langle x \rangle\overset{\text{def}}{=}(x)x(u).(\overline{x}u\mid Diverge\langle x\rangle)$. We see that all messages are delivered eventually. 

\section{A Theory of True Concurrency for $\textrm{A}\pi_{tc}$}\label{ttca}

\subsection{True Concurrency}\label{tc}

Firstly, in this subsection, the related concepts on true concurrency are defined based on the following concepts \cite{ES1} \cite{ES2} \cite{CM}.

\begin{definition}[Prime event structure]\label{PES}
Let $\Lambda$ be a fixed set of labels, ranged over $a,b,c,\cdots$. A ($\Lambda$-labelled) prime event structure is a tuple $\mathcal{E}=\langle \mathbb{E}, \leq, \sharp, \lambda\rangle$, where $\mathbb{E}$ is a denumerable set of events. Let $\lambda:\mathbb{E}\rightarrow\Lambda$ be a labelling function. And $\leq$, $\sharp$ are binary relations on $\mathbb{E}$, called causality and conflict respectively, such that:

\begin{enumerate}
  \item $\leq$ is a partial order and $\lceil e \rceil = \{e'\in \mathbb{E}|e'\leq e\}$ is finite for all $e\in \mathbb{E}$.
  \item $\sharp$ is irreflexive, symmetric and hereditary with respect to $\leq$, that is, for all $e,e',e''\in \mathbb{E}$, if $e\sharp e'\leq e''$, then $e\sharp e''$.
\end{enumerate}

Then, the concepts of consistency and concurrency can be drawn from the above definition:

\begin{enumerate}
  \item $e,e'\in \mathbb{E}$ are consistent, denoted as $e\frown e'$, if $\neg(e\sharp e')$. A subset $X\subseteq \mathbb{E}$ is called consistent, if $e\frown e'$ for all $e,e'\in X$.
  \item $e,e'\in \mathbb{E}$ are concurrent, denoted as $e\parallel e'$, if $\neg(e\leq e')$, $\neg(e'\leq e)$, and $\neg(e\sharp e')$.
\end{enumerate}
\end{definition}

\begin{definition}[Configuration]
Let $\mathcal{E}$ be a PES. A (finite) configuration in $\mathcal{E}$ is a (finite) consistent subset of events $C\subseteq \mathcal{E}$, closed with respect to causality (i.e. $\lceil C\rceil=C$). The set of finite configurations of $\mathcal{E}$ is denoted by $\mathcal{C}(\mathcal{E})$.
\end{definition}

Usually, truly concurrent behavioral equivalences are defined by events $e\in\mathcal{E}$ and prime event structure $\mathcal{E}$, in contrast to interleaving behavioral equivalences by actions $a,b\in\mathcal{P}$ and process (graph) $\mathcal{P}$. Indeed, they have correspondences, in \cite{SFL}, models of concurrency, including Petri nets, transition systems and event structures, are unified in a uniform representation -- TSI (Transition System with Independence).

If $x$ is a process, let $C(x)$ denote the corresponding configuration (the already executed part of the process $x$, of course, it is free of conflicts), when $x\xrightarrow{e} x'$, the corresponding configuration $C(x)\xrightarrow{e}C(x')$ with $C(x')=C(x)\cup\{e\}$, where $e$ may be caused by some events in $C(x)$ and concurrent with the other events in $C(x)$, or entirely concurrent with all events in $C(x)$, or entirely caused by all events in $C(x)$. With a little abuse of concepts, in the following of the paper, we will not distinguish actions and events, prime event structures and processes, also concurrent behavior equivalences based on configurations and processes, and use them freely, unless they have specific meanings.

Next, we introduce concepts of truly concurrent bisimilarities, including pomset bisimilarity, step bisimilarity, history-preserving (hp-)bisimilarity and hereditary history-preserving (hhp-)bisimilarity. In contrast to traditional truly concurrent bisimilarities in CTC \cite{CTC} and APTC \cite{ATC}, these versions in $\textrm{A}\pi_{tc}$ not only must take care of actions with bound objects, but also must suit for the constraints of the type system. That is, the truly concurrent bisimilarities are tagged with a parameter $\rho$. Note that, here, a PES $\mathcal{E}$ is deemed as a configuration.

\begin{definition}[Pomset transitions and step]
Let $\mathcal{E}$ be a PES and let $C\in\mathcal{C}(\mathcal{E})$, and $\emptyset\neq X\subseteq \mathbb{E}$, if $C\cap X=\emptyset$ and $C'=C\cup X\in\mathcal{C}(\mathcal{E})$, then $C\xrightarrow{X} C'$ is called a pomset transition from $C$ to $C'$. When the events in $X$ are pairwise concurrent, we say that $C\xrightarrow{X}C'$ is a step.
\end{definition}

\begin{definition}[Pomset, step bisimilarity]\label{PSB}
Let $\mathcal{E}_1$, $\mathcal{E}_2$ be PESs. A pomset bisimulation is a relation $R\subseteq\mathcal{C}(\mathcal{E}_1)\times\mathcal{C}(\mathcal{E}_2)$, such that if $(C_1,C_2)\in R$ with $rcp(C_1)=rcp(C_2)$, and $C_1\xrightarrow{X_1}C_1'$ (with $\mathcal{E}_1\xrightarrow{X_1}\mathcal{E}_1'$) then $C_2\xrightarrow{X_2}C_2'$ (with $\mathcal{E}_2\xrightarrow{X_2}\mathcal{E}_2'$), with $X_1\subseteq \mathbb{E}_1$, $X_2\subseteq \mathbb{E}_2$, $X_1\sim X_2$, $rcp(C_1')=rcp(C_2')$ and $(C_1',C_2')\in R$:
\begin{enumerate}
  \item for each fresh action $\alpha\in X_1$, if $C_1''\xrightarrow{\alpha}C_1'''$ (with $\mathcal{E}_1''\xrightarrow{\alpha}\mathcal{E}_1'''$), then for some $C_2''$ and $C_2'''$, $C_2''\xrightarrow{\alpha}C_2'''$ (with $\mathcal{E}_2''\xrightarrow{\alpha}\mathcal{E}_2'''$), such that if $(C_1'',C_2'')\in R$ with $rcp(C_1'')=rcp(C_2'')$ then $(C_1''',C_2''')\in R$ with $rcp(C_1''')=rcp(C_2''')$;
  \item for each $x(y)\in X_1$ with ($y\notin n(\mathcal{E}_1, \mathcal{E}_2)$), if $C_1''\xrightarrow{x(y)}C_1'''$ (with $\mathcal{E}_1''\xrightarrow{x(y)}\mathcal{E}_1'''\{w/y\}$) for all $w$, then for some $C_2''$ and $C_2'''$, $C_2''\xrightarrow{x(y)}C_2'''$ (with $\mathcal{E}_2''\xrightarrow{x(y)}\mathcal{E}_2'''\{w/y\}$) for all $w$, such that if $(C_1'',C_2'')\in R$ with $rcp(C_1'')=rcp(C_2'')$ then $(C_1''',C_2''')\in R$ with $rcp(C_1''')=rcp(C_2''')$;
  \item for each two $x_1(y),x_2(y)\in X_1$ with ($y\notin n(\mathcal{E}_1, \mathcal{E}_2)$), if $C_1''\xrightarrow{\{x_1(y),x_2(y)\}}C_1'''$ (with $\mathcal{E}_1''\xrightarrow{\{x_1(y),x_2(y)\}}\mathcal{E}_1'''\{w/y\}$) for all $w$, then for some $C_2''$ and $C_2'''$, $C_2''\xrightarrow{\{x_1(y),x_2(y)\}}C_2'''$ (with $\mathcal{E}_2''\xrightarrow{\{x_1(y),x_2(y)\}}\mathcal{E}_2'''\{w/y\}$) for all $w$, such that if $(C_1'',C_2'')\in R$ with $rcp(C_1'')=rcp(C_2'')$ then $(C_1''',C_2''')\in R$ with $rcp(C_1''')=rcp(C_2''')$;
  \item for each $\overline{x}(y)\in X_1$ with $y\notin n(\mathcal{E}_1, \mathcal{E}_2)$ and $x\notin rcp(C_1'')\cup rcp(C_2'')$, if $C_1''\xrightarrow{\overline{x}(y)}C_1'''$ (with $\mathcal{E}_1''\xrightarrow{\overline{x}(y)}\mathcal{E}_1'''$), then for some $C_2''$ and $C_2'''$, $C_2''\xrightarrow{\overline{x}(y)}C_2'''$ (with $\mathcal{E}_2''\xrightarrow{\overline{x}(y)}\mathcal{E}_2'''$), such that if $(C_1'',C_2'')\in R$ with $rcp(C_1'')=rcp(C_2'')$ then $(C_1''',C_2''')\in R$ with $rcp(C_1''')=rcp(C_2''')$;
  \item for each $\overline{x}(y)\in X_1$ with $y\notin n(\mathcal{E}_1, \mathcal{E}_2)$ and $x\in rcp(C_1'')=rcp(C_2'')$, if $C_1''\xrightarrow{\overline{x}(y)}C_1'''$ (with $\mathcal{E}_1''\xrightarrow{\overline{x}(y)}\mathcal{E}_1'''$), then for some $C_2''$ and $C_2'''$, either $C_2''\xrightarrow{\overline{x}(y)}C_2'''$ (with $\mathcal{E}_2''\xrightarrow{\overline{x}(y)}\mathcal{E}_2'''$), or $C_2''\Rightarrow C_2'''$ (with $\mathcal{E}_2''\Rightarrow\mathcal{E}_2'''$), such that if $(C_1'',C_2'')\in R$ with $rcp(C_1'')=rcp(C_2'')$ then $(C_1''',C_2''')\in R$ with $rcp(C_1''')=rcp(C_2''')$.
\end{enumerate}
 and vice-versa.

We say that $\mathcal{E}_1$, $\mathcal{E}_2$ are pomset bisimilar, written $\mathcal{E}_1\sim_p\mathcal{E}_2$, if there exists a pomset bisimulation $R$, such that $(\emptyset,\emptyset)\in R$. By replacing pomset transitions with steps, we can get the definition of step bisimulation. When PESs $\mathcal{E}_1$ and $\mathcal{E}_2$ are step bisimilar, we write $\mathcal{E}_1\sim_s\mathcal{E}_2$.
\end{definition}

\begin{definition}[Posetal product]
Given two PESs $\mathcal{E}_1$, $\mathcal{E}_2$, the posetal product of their configurations, denoted $\mathcal{C}(\mathcal{E}_1)\overline{\times}\mathcal{C}(\mathcal{E}_2)$, is defined as

$$\{(C_1,f,C_2)|C_1\in\mathcal{C}(\mathcal{E}_1),C_2\in\mathcal{C}(\mathcal{E}_2),f:C_1\rightarrow C_2 \textrm{ isomorphism}\}.$$

A subset $R\subseteq\mathcal{C}(\mathcal{E}_1)\overline{\times}\mathcal{C}(\mathcal{E}_2)$ is called a posetal relation. We say that $R$ is downward closed when for any $(C_1,f,C_2),(C_1',f',C_2')\in \mathcal{C}(\mathcal{E}_1)\overline{\times}\mathcal{C}(\mathcal{E}_2)$, if $(C_1,f,C_2)\subseteq (C_1',f',C_2')$ pointwise and $(C_1',f',C_2')\in R$, then $(C_1,f,C_2)\in R$.

For $f:X_1\rightarrow X_2$, we define $f[x_1\mapsto x_2]:X_1\cup\{x_1\}\rightarrow X_2\cup\{x_2\}$, $z\in X_1\cup\{x_1\}$,(1)$f[x_1\mapsto x_2](z)=
x_2$,if $z=x_1$;(2)$f[x_1\mapsto x_2](z)=f(z)$, otherwise. Where $X_1\subseteq \mathbb{E}_1$, $X_2\subseteq \mathbb{E}_2$, $x_1\in \mathbb{E}_1$, $x_2\in \mathbb{E}_2$.
\end{definition}

\begin{definition}[(Hereditary) history-preserving bisimilarity]\label{HHPB}
A history-preserving (hp-) bisimulation is a posetal relation $R\subseteq\mathcal{C}(\mathcal{E}_1)\overline{\times}\mathcal{C}(\mathcal{E}_2)$ such that if $(C_1,f,C_2)\in R$ with $rcp(C_1)=rcp(C_2)$, and
\begin{enumerate}
  \item for $e_1=\alpha$ a fresh action, if $C_1\xrightarrow{\alpha}C_1'$ (with $\mathcal{E}_1\xrightarrow{\alpha}\mathcal{E}_1'$), then for some $C_2'$ and $e_2=\alpha$, $C_2\xrightarrow{\alpha}C_2'$ (with $\mathcal{E}_2\xrightarrow{\alpha}\mathcal{E}_2'$), such that $(C_1',f[e_1\mapsto e_2],C_2')\in R$ with $rcp(c_1')=rcp(C_2')$;
  \item for $e_1=x(y)$ with ($y\notin n(\mathcal{E}_1, \mathcal{E}_2)$), if $C_1\xrightarrow{x(y)}C_1'$ (with $\mathcal{E}_1\xrightarrow{x(y)}\mathcal{E}_1'\{w/y\}$) for all $w$, then for some $C_2'$ and $e_2=x(y)$, $C_2\xrightarrow{x(y)}C_2'$ (with $\mathcal{E}_2\xrightarrow{x(y)}\mathcal{E}_2'\{w/y\}$) for all $w$, such that $(C_1',f[e_1\mapsto e_2],C_2')\in R$ with $rcp(C_1')=rcp(C_2')$;
  \item for $e_1=\overline{x}(y)$ with $y\notin n(\mathcal{E}_1, \mathcal{E}_2)$ and $x\notin rcp(C_1'')\cup rcp(C_2'')$, if $C_1\xrightarrow{\overline{x}(y)}C_1'$ (with $\mathcal{E}_1\xrightarrow{\overline{x}(y)}\mathcal{E}_1'$), then for some $C_2'$ and $e_2=\overline{x}(y)$, $C_2\xrightarrow{\overline{x}(y)}C_2'$ (with $\mathcal{E}_2\xrightarrow{\overline{x}(y)}\mathcal{E}_2'$), such that $(C_1',f[e_1\mapsto e_2],C_2')\in R$ with $rcp(C_1')=rcp(C_2')$;
  \item for $e_1=\overline{x}(y)$ with $y\notin n(\mathcal{E}_1, \mathcal{E}_2)$ and $x\in rcp(C_1'')=rcp(C_2'')$, if $C_1\xrightarrow{\overline{x}(y)}C_1'$ (with $\mathcal{E}_1\xrightarrow{\overline{x}(y)}\mathcal{E}_1'$), then for some $C_2'$ and $e_2=\overline{x}(y)$, either $C_2\xrightarrow{\overline{x}(y)}C_2'$ (with $\mathcal{E}_2\xrightarrow{\overline{x}(y)}\mathcal{E}_2'$), such that $(C_1',f[e_1\mapsto e_2],C_2')\in R$ with $rcp(C_1')=rcp(C_2')$; or $C_2\Rightarrow C_2'$ (with $\mathcal{E}_2\Rightarrow\mathcal{E}_2'$), such that $(C_1',f[e_1\mapsto \tau],C_2')\in R$ with $rcp(C_1')=rcp(C_2')$;
\end{enumerate}

and vice-versa. $\mathcal{E}_1,\mathcal{E}_2$ are history-preserving (hp-)bisimilar and are written $\mathcal{E}_1\sim_{hp}\mathcal{E}_2$ if there exists a hp-bisimulation $R$ such that $(\emptyset,\emptyset,\emptyset)\in R$.

A hereditary history-preserving (hhp-)bisimulation is a downward closed hp-bisimulation. $\mathcal{E}_1,\mathcal{E}_2$ are hereditary history-preserving (hhp-)bisimilar and are written $\mathcal{E}_1\sim_{hhp}\mathcal{E}_2$.
\end{definition}

Since the Parallel composition $\mid$ is a fundamental computational pattern in CTC, APTC and $\pi_{tc}$, and also it is fundamental in $\textrm{A}\pi_{tc}$ as defined in Table \ref{TR}, and cannot be instead of other operators.

\subsection{Algebraic Laws}

Similarly, for an index set $I=\{1,\cdots,n\}$, we use $\sum_{i\in I}P_i$ to denote $(\nu u)(\textrm{case }u \textrm{ of }(u:P_1,\cdots,u:P_n))$ for $u$ fresh if $I\neq\emptyset$; $\textbf{0}$ otherwise. If $I$ is a singleton, we write $\sum P$ instead of $\sum_{i\in I}P$. And we also let the variable $G$ range over processes $\sum_{i\in I}P_i$. Then we get the following axioms as Table \ref{AL} shows.

\begin{center}
    \begin{table}
        \[\textrm{A1}\quad G+G=G \quad \textrm{A2}\quad G+\textbf{0}=G\]

        \[\textrm{A3}\quad P\mid\textbf{0}=P \quad \textrm{A4}\quad P\mid Q=Q\mid P \quad \textrm{A5}\quad (P\mid Q)\mid R=P\mid(Q\mid R)\]

        \[\textrm{A6}\quad (\nu x)(\sum_{i\in I}P_i)=\sum_{i\in I}(\nu x)P_i\]

        \[\textrm{A7}\quad (\nu x)(P\mid Q)=P\mid (\nu x)Q\quad (x\notin n(P))\]

        \[\textrm{A8}\quad (\nu x)(\overline{x}y\mid\alpha.P)=\alpha.(\nu x)(\overline{x}y\mid P)\quad (x\notin n(\alpha))\]

        \[\textrm{A9}\quad (\nu x)(\overline{x}y\mid x(z).P)=(\nu x)(P\{y/z\})\]

        \[\textrm{A10}\quad (\nu x)(y(z).P)=y(z).(\nu x)P \quad (x\neq y, x\neq z)\]

        \[\textrm{A11}\quad \overline{x}y\mid \sum_{i\in I}P_i=\sum_{i\in I}(\overline{x}y\mid P_i) \quad (I\neq\emptyset)\]

        \[\textrm{A12}\quad \alpha.\sum_{i\in I}P_i=\sum_{i\in I}\alpha.P_i \quad (I\neq\emptyset)\]

        \[\textrm{A13}\quad P=\sum P\]

        \[\textrm{A14}\quad \overline{u}v\mid(x(y).P)=(\overline{u}v\mid x(y)).P \quad (y\neq u, y\neq v)\]

        \[\textrm{A15}\quad (\overline{u}v.P)\mid x(y)=(\overline{u}v\mid x(y)).P \quad (y\neq u, y\neq v)\]

        \[\textrm{A16}\quad (\overline{u}v).P\mid(x(y).Q)=(\overline{u}v\mid x(y)).(P\mid Q) \quad (y\neq u, y\neq v)\]

        \[\textrm{A17}\quad \overline{x}y\mid(x(y).P)=\tau.P\]

        \[\textrm{A18}\quad (\overline{x}y.P)\mid x(y)=\tau.P\]

        \[\textrm{A19}\quad (\overline{x}y).P\mid(x(y).Q)=\tau.(P\mid Q)\]

        \[\textrm{A20}\quad \overline{x}y\mid (z(w).P)=\sum(\overline{x}y\mid z(w)).P + \sum z(w).P + \sum Q\quad (x\in rcp(\overline{x}y\mid (z(w).P)), w\neq x, w\neq y, Q=P\{y/w\}\textrm{ if }x=z; Q=\textbf{0}\textrm{ otherwise})\]
        
        \caption{Algebraic laws of $\textrm{A}\pi_{tc}$}
        \label{AL}
    \end{table}
\end{center}

Then we have the following conclusions.

\begin{theorem}[Soundness modulo pomset bisimilarity]
The axioms in Table \ref{AL} are sound modulo pomset bisimilarity.
\end{theorem}

\begin{theorem}[Soundness modulo step bisimilarity]
The axioms in Table \ref{AL} are sound modulo step bisimilarity.
\end{theorem}

\begin{theorem}[Soundness modulo hp-bisimilarity]
The axioms in Table \ref{AL} are sound modulo hp-bisimilarity.
\end{theorem}

\begin{theorem}[Soundness modulo hhp-bisimilarity]
The axioms in Table \ref{AL} are sound modulo hhp-bisimilarity.
\end{theorem}

\section{Conclusions}\label{con}

Based on our previous work on process algebra for true concurrency CTC \cite{CTC}, APTC \cite{ATC} and $\pi_{tc}$ \cite{PITC}, we adjust $\textrm{A}\pi$ \cite{Actor4} to make it have a truly concurrent semantics. Since the actor computational model is a model for true concurrency, $\textrm{A}\pi_{tc}$ makes the algebra of actors \emph{truly} true concurrency.

\label{lastpage}

\end{document}